# A Combined Environmental Monitoring Framework based on WSN Clustering and VANET Edge Computation Offloading

Basilis Mamalis
University of West Attica
Ag. Spyridonos, 12243,
Athens, Greece
vmamalis@uniwa.gr

Sergios Gerakidis
Hellenic Open University
Aristotelous 18, 26335
Patras, Greece
sgerakidis@gmail.com

## ABSTRACT
Wireless Sensor Networks (WSN) and Vehicular Ad-hoc Networks (VANET) have been extensively used in IoT applications for environmental monitoring, especially in rural and agricultural areas. In this paper we present a novel combined approach which uses both WSN and VANET clustered structures for the efficient gathering and processing of environmental parameters in long eNodeB/RSU-enabled roads/highways, which cross large rural areas. The former (WSNs) is used for traditional sensing and data gathering, whereas the latter (VANET) is used for both (a) performing intermediate processing based on modern edge computation offloading techniques, and (b) propagating the data (either raw data or computed results) to the residing eNodeB/RSUs. Extended experimental measurements, taken via the combined use of SUMO and Veins simulation platforms (and focusing in the air quality monitoring experimental case), demonstrate the high efficiency and scalability of the presented approach over very large deployment areas.

## General Terms
Networks, Sensors, Vehicles, Algorithms

## Keywords
Wireless Sensor Network; Vehicular Ad-hoc Network; Node Clustering; Network Lifetime; Virtual Edge Computing, Road Side Unit; Computation Offloading

## 1. INTRODUCTION
The industrial growth, the rapid technological evolution, and the increased human activities for comfort and sumptuous living, affect significantly several environmental parameters like humidity, temperature, methane, carbon monoxide, smoke etc. The continuous increase of these parameter values has many negative effects on the health as well as on the daily life of the people, and it's also regarded as the main reason for several diseases, like cancers etc.

As a consequence, it's nowadays essential to monitor the environmental parameters of a region frequently. Wireless sensor networks (WSNs) can efficiently monitor various environmental conditions, including the following: pressure, sound, motion, temperature, vibration, acceleration, humidity, and also pollutant/chemical concentrations [1,2]. Considering deployments under such conditions, the goal of a WSN is to support the fast propagation of the sensed data (probably through their self-organization in clusters and executing a relevant cluster-based routing protocol) to a fixed or mobile base station, where it can be processed and analyzed [3].

The deployment of the sensors of a WSN can be static or dynamic; in either case an extended geographical area is potentially covered in case of environmental monitoring applications [4]. However, several important issues usually arise in large WSN deployments, that may influence the quality of service significantly. As an example, with regard to deployments in dynamic environments, the gathered data may be affected due to several weather changes or human activities [2,5]. As a consequence, the need for real-time monitoring of crucial environmental parameters is obvious. In such cases, the collaboration with other networks and resources (such as VANETs or FANETs) may also be a good alternative.

Conventional systems, in order to scan/monitor the air, soil and water, mostly use filtration, absorption, electrostatic sampling, condensation, sedimentation etc. On the other hand, the daily activity of humans and animals often cause damage in the environmental resources, affecting people's quality of life. Wireless sensors have been extensively used in the last decade by researchers, to collect as much as possible, on-time, continuous, accurate data for crucial parameters, suitable for high-quality/reliability analysis [4,6]. For example, common sensors are used for direct air or soil quality monitoring, in order to detect hazardous substances, and then examine it in more details for possible threats in human health [7].

Moreover, with the use of sensors we can reliably realize the quality of water, for example by measuring and analyzing several biological, chemical, and physical parameters [8-10]. Through the use of WSNs, the relevant procedures tend to be simpler and even more reliable and practical. However, several important challenges in the design and implementation of WSNs still exist, including the coverage of the network, the energy consumption, the real-time response needs of modern applications, and the increased security needs [11].

The automated behavior of environmental applications nowadays, also offers the capability of collecting, transmitting and storing large portions of data into a global datastore, suitable for further big data processing and analysis. This capability guarantees (through proper mid-term and long-term actions) the conservation of high-quality standards in air and water quality in big cities [10,11]. Furthermore, the rapid evolution of smart cities, enhanced by the green computing paradigm, is one of the main reasons behind the increased trend for development of WSN systems for environmental monitoring applications. As a consequence, it's crucial to efficiently support all the basic operations of the underlying WSN, to offer the required services, and fulfill the design objectives of the relevant application [12].

So, the use of WSNs in almost all kinds of environmental monitoring applications has been very popular during the last decade, since automation and new technologies are able to offer real-time (or near real-time) monitoring, and generally, faster (more intelligent and more accurate) overall solutions. The most modern integrated WSN-based solutions for environmental monitoring applications include the efficient handling of all the involved procedures, like collecting, processing, transferring, storing, retrieving etc. Real-time data access, short-term monitoring actions, long-term monitoring actions and further processing, as well as increased scalability, are usually provided by such WSN solutions [13]. The WSNs for environmental monitoring applications are mainly ad-hoc, multi-hop structures with dynamic behavior. Moreover, the architecture, the scale and the complexity of the required WSN, is often affected by the kind of the specific application. Also, WSNs for such applications usually focus on efficient data gathering and energy conservation [14].

Moreover, during the last decade, VANETs have been broadly followed as a proper solution for measuring and processing the environmental conditions in corresponding IoT applications, mainly focusing in agricultural as well as rural target fields. Specifically, VANETs can been efficiently used for both sensing and direct transferring the sensed values collected by the SNs deployed around. As a good example, in the work of [15], the authors present the use of VANET as an ideal solution for monitoring environmental conditions in any place, because of the ad-hoc structure they introduce, which offers increased flexibility and adequate coverage in large / very large / huge target spaces.

Automated monitoring of the environmental conditions in any place involves, among else, the below actions: measuring the relevant parameter values in every current position, forwarding them from the current position to the center of observation, and finally, their further processing and analysis using the center's resources. Beyond this, various important parameters of the underlying network are estimated and analyzed, like the overall latency, the total count of hops, the caused gaps etc. The above parameters are calculated with respect to several kinds of measured environmental conditions, and are then being compared in terms of various well-known VANET protocols for routing. These protocols include the GSR (geographic source routing) protocol, the A-STAR (anchor-based street and traffic-aware routing) protocol, and the P-GEDIR (peripheral node based geographic distance routing) protocol. The motion patterns of the vehicles are formed with the use of SUMO (simulation of urban mobility) simulation tool, whereas the relevant network factors (overall latency, hops count etc.), are extracted with the use of OMNET++ network simulator.

In [16,17], the reader may also find some additional applications with similar objectives, which include automated collecting and processing of sensed parameter values of the environment with VANET support. Furthermore, in [18], an automated VANET-based framework for real-time monitoring of fire condition in a rural area is presented.

The transmission of the measured parameter values, either from the place of measurement to the center of observation, or from one position to another is usually performed through vehicular cloud. In [19], the authors present the integration of VANET data mining services and IoT techniques in a single multilayer cloud-oriented platform. Furthermore, considering a vehicle with modern equipment, the integration of many services, preserving high quality for all of them, is not an easy task. With respect to the specific requirements and the necessary service policies in such situations, the authors in [20] propose a relevant, notably efficient technique, for 5G-enabled vehicular clouds. In a similar way, the authors in [21] elaborate the integration of mobile/cloud computing, VANET, and ITS techniques, to construct a smart system for disaster management. Furthermore, the authors in [22] present a combined solution involving VANET and cloud computing support, aiming at the efficient management of the resources of a modern vehicle (storage, computational, and bandwidth resources). So, the technological evolution in ITS and VANETs has led to the extensive use of the vehicular cloud, for various applications with real-time needs. Two other important issues in VANET applications are the determination of neighbor vehicles positions [23], as well as the selection of the routing protocol [25]. In order to determine the position of neighboring vehicles, VANET use the broadcast protocol. But some limitations exist (mainly depending on the network density), like contention, broadcast storm, and failure in message delivery. Several message dissemination protocols have been proposed in the literature, whose main objective is to face the above limitations [23,24]. With respect to the routing protocols used in VANET applications, due to the variable speeds of vehicles, the frequent link disconnections, and the constrained motion of the vehicles around the road, the position-based or geographic routing protocols are generally preferred [25]. A detailed overview in VANET routing protocols for IoT applications can be found in [26,27].

Trying to exploit the advantages of both types of ad-hoc networks (WSNs, VANETs) reviewed above with regard to their extensive use in environmental monitoring applications, and achieve faster data gathering and processing of crucial parameters, we've designed a novel combined protocol, which is strongly suitable for direct application on long RSU-enabled roads/highways, which cross large rural areas. The WSN clustered structure is used for traditional sensing and multi-hop data gathering, whereas the underlying VANET platform is appropriately used for fast propagating to the residing RSUs, performing also intermediate processing through modern computation offloading techniques. The latter is based on the suitable formation of stable vehicle clusters, which enhance the capability of finding one or more virtual edge computing nodes (cluster members), appropriate for real-time task offloading within each cluster. A corresponding simulation experiment is also conducted, using SUMO, Veins and Castalia simulation platforms, in order to demonstrate the high efficiency of the proposed approach in the experimental field of air quality monitoring.

The remaining text is organized as follows. In section 2, the necessary background is given with respect to the WSN and VANET clustering protocols used as the basis of the proposed combined approach. Then, in section 3, our compound scheme for environmental monitoring, elaborating both WSN and VANET ad-hoc structures in proper collaboration, is presented along with relevant discussion. In section 4, a complete set of simulated experiments are presented and the corresponding results are thoroughly discussed, whereas section 5 concludes the paper.

## 2. MATERIALS AND METHODS

### 2.1 The WSN clustering approach

With regard to the cluster-based organization of the sensor nodes, we've used a carefully designed protocol, based on the main idea presented in [28]. The relevant cluster formation protocol has been selected as the most suitable for the purposes of the proposed combined edge computation scheme due the following advantages:

- It leads to highly energy-balanced clusters, keeping in parallel very fast execution time. Note here, that we don't use a more complicated protocol (which would probably behave slightly better in energy balancing), in order to preserve the fastest possible execution and maintenance time, even in very large rural areas.

- It can be easily adjusted (keeping all of its efficiency as well as its energy-balancing behavior) to non-symmetric, large-scale deployment areas, with different base station positions, due to the capability of appropriately setting the relevant value of the forwarding angle.

More concretely, our underlying multi-hop clustering scheme follows as its main cluster formation criterion the 'remaining energy' of each SN, thus finally leading to energy-balanced clusters, and also to efficient handling of the 'energy holes' formed around the CHs. So, it's naturally expected to prolong the network lifetime. In more details, the cluster formation algorithm consists of the following steps:

a. In the beginning, all the nodes of the WSN broadcast messages (including their remaining energy and their identity) in a fixed power, which guarantees that nodes within a radius R (which is a pre-specified threshold) will receive the message. Then, each node waits to receive relevant messages from all its 'neighboring' nodes.

b. For each such received message, the recipient node first determines if the message was sent from a SN lying in the 'direction' from the node to the BS. The necessary check is based on the value of angle ($\theta$) formed among the lines connecting the node that received the message to the sender node and to the BS respectively.

c. If that holds, it then compares the remaining energy referred in the message with its own, and acts as follows: If the energy referred in the message is larger, it specifies the node which sent the message as its *parent*.

d. If the recipient node has already specified another node as its parent, and the node which sent the message refers larger remaining energy than its own, then it checks if the distance between itself and its parent node is larger than the distance between itself and the node which sent the message. If that holds, it replaces its parent node with the node who sent the message.

e. Upon a node has received the relevant messages from all its neighboring nodes, and it has made the necessary decisions, it sends a 'join' message to its own parent, and specifies itself as a 'member' node.

f. If the node has no parent node, then it specifies itself as a clusterhead and broadcasts a corresponding message.

As a result, at the end of the execution of the above protocol, every node has specified as its 'parent' the node that has greater remaining energy than itself, with the minimum distance. The relevant simulation experiments in [28] indicate that the above protocol reduces the energy costs significantly, achieves balanced energy consumption, and maximizes the network lifetime. Further, the most important goal of the protocol is the formation of proper (energy-balanced) clusters with not only high-energy CHs, but also with energy-rich neighborhoods. So, it can easily handle the energy holes around the clusterheads and sufficiently support the energy-efficient data gathering in large scale deployment areas.

It should also be noted here, that considering the inter-cluster communication, each clusterhead sends the received data (either the data coming from nodes of other clusterheads, or data forwarded by the member-nodes of its cluster), in a round-robin basis, to neighboring nodes of other clusters lying within the 'direction' from the clusterhead to the BS (based on angle $\theta$). More details can be found in [28].

### 2.2 The VANET clustering structure and virtual edge formation

The VANET-based edge computation offloading scheme of our combined processing framework is supported on its basis by the clustering protocol presented in [29]. The total approach is firstly based on the creation of stable clusters, whose members are then regarded as good candidate VEC providers (i.e. servers for task offloading). More specifically, through the above clustering protocol, appropriate multi-hop clusters are formed, with not only stable clusterheads, but also with highly stable neighborhoods. This feature is inherited recursively till the nodes at the leaves of the cluster. Also, a suitable RSU/MEC-enabled maintenance mechanism has been carefully designed and implemented to further increase the stability of the involved clusters.

Additionally, a properly documented RSU/MEC-enabled trust management mechanism is introduced to support the resistance of the whole network to specific kinds of security attacks. In brief, the proposed algorithm for clusters formation consists of the steps below:

i. Initially, every vehicle *i* calculates its *Mobility Difference* (*MD value*) with every neighbor *j*, and forms a set *U* of probable neighbors, the vehicles for which the calculated *MD* value doesn't overdraw a fixed threshold (*Ts*) and follow the same direction. Based on the above calculation, it determines the most suitable neighbors (i.e. the vehicles with the most similar mobility behavior), which then will be the basic candidates to choose one to follow as its *parent* node. The *MD* value for every neighbor vehicle *j* with regard to *i* is calculated taking in account the *speed* and *acceleration differences* ($SD(i,j)$, $AD(i,j)$ respectively – see [29] for the detailed definition) among the two vehicles, which can be regarded as the most crucial factors giving the actual similarity in the mobility patterns of the two vehicles. The actual *MD* value is finally given by the following expression, where $c_1$ and $c_2$ are suitably selected constant coefficients (with $c_1 + c_2 = 1$):

$$MD(i,j) = c_1 \cdot SD(i,j) + c_2 \cdot AD(i,j)$$

ii. Every vehicle *i* also calculates its total Stability Factor (*SF*) value. The *SF* value of a vehicle is calculated taking in account the overall/average differences in speed and acceleration, the degree of the vehicle, as well as its overall/average distance from all its neighbors. More specifically, it's calculated as follows:

$$SF(i) = \alpha \cdot SD_{av}(i) + \beta \cdot AD_{av}(i) + \gamma \cdot D_{av}(i) + \delta \cdot d(i)$$

In this calculation, the coefficients *α, β, γ* and *δ* are properly selected in such a way that $\alpha + \beta + \gamma + \delta = 1$, *d(i)* represents the degree of node *i*, and $SD_{av}(i)$, $AD_{av}(i)$, $D_{av}(i)$ are the relevant measurements of the average speed difference, average acceleration difference and average relative distance, respectively, among vehicle *i* and all its neighboring vehicles (see [29] for more details).

iii. Taking in account the SF values advertised by all its neighboring vehicles, every vehicle *i* selects to follow as its parent the most appropriate neighbor node (among the candidates that are eligible). More specifically, *i* selects as its *parent* node, the node with the maximum total Stability Factor (*SF*), given that the corresponding value is higher than its own SF value too.

iv. Eventually, the vehicle with no neighboring node that satisfies both the necessary conditions (sufficient small *MD* value – i.e. mobility similarity, and higher *SF* value), marks itself as a *clusterhead* (*CH*).

As a result, at the end of the execution of the above protocol, each vehicle has selected as its *parent* a vehicle that has sufficiently similar mobility pattern (behavior) and the maximum overall stability among its neighbors. Based on this feature, which progressively extends till the root node (which is selected as the head of the cluster), the proposed algorithm naturally leads to the formation of highly stable clusters with increased lifetime. In [29], the reader may also find specific details with respect to the mechanism used to maintain clusters' stability, as well as with respect to trust management.

Upon its formation, the above stable cluster organization, can also serve as a virtual edge for reliable computation offloading. Specifically, when one or more tasks are generated on a vehicle/member of a cluster, another (the most suitable) cluster member (CM) can be efficiently found to act as a virtual edge computing (VEC) server for offloading the requested task/s, as proposed in [30].

## 3. THE PROPOSED COMBINED EDGE COMPUTATION SCHEME

The proposed WSN/VANET combined edge computation scheme, first assumes that we are given a large rural area which is crossed by a highway, in which it is necessary to take environmental measurements and process them in real or near real time. In this direction, the main objective of the designed framework is to combine and exploit the advantages offered by VANETs and WSNs, in order to provide an efficient solution that can be extended to very large scale. The details of the proposed approach are given in the rest of the section.

### *A. VANET Setup and Assumptions:*

With respect to the connected vehicles support, the proposed approach is based on the following assumptions (similar to the ones followed in [2,3]). First, we assume (without loss of generality, see [3]) that we have a three-lanes road/highway in both directions, with eNodeB/RSU and MEC server support every X kilometers. Second, each vehicle moving in the road has a unique *ID* in the network. Third, every vehicle is equipped with digital road map and a GPS device that allows it to obtain its real-time geolocation and knows its speed and maximal acceleration. Fourth, each vehicle is equipped with wireless interface, in order to communicate with other vehicles, eNodeB/RSUs and RNs. Finally, all the vehicles in the road/highway form sufficiently stable clusters according to the algorithm described in section 2.2.

### *B. WSN Setup and Assumptions:*

A suitable number of SNs are deployed at random, covering the area we are interested for environmental measurements; typically, a large area surrounding the main road/highway, in both sides. This area could be as large as is in tis physical extension (in both sides), provided that the RSU/MEC-enabled main road/highway continuous existing in the middle. The sensor nodes are then self-organized in a number of energy-balanced clusters, following the cluster formation algorithm described in section 2.1, and also adopting some necessary changes as given below:

a. A set of 'virtual' base-station points are defined in tactical intervals between the RSUs residing in the road. More concretely, one such 'virtual' base-station point is positioned every Y meters among two adjacent RSUs.

b. Each one of the above virtual base-station points corresponds to a relevant SNs deployment sub-area. The extend of each such sub-area is specified by the limits +Y/2 and -Y/2 meters (left and right respectively) from each virtual base-station point.

c. So, all the SNs are then assumed to execute the cluster formation algorithm of section 2.1, in their deployment area (i.e. realizing as their base station, the corresponding virtual base-station point defined above), by suitably setting the value of angle *θ* according to the location of their own virtual base-station point.

In that way, after the end of execution of the cluster formation algorithm, there will be a sufficient number of cluster-heads (CHs), residing very close to the road/highway (in both sides). These CHs will be called in the rest of the paper as the rendezvous nodes (RNs) of the underlying WSN structure, and they will be responsible for the whole communication with the vehicles coming across the road/highway. Note also here that, with respect to the values of the parameters X and Y referred above, X may range from 3 to 5 kilometers, and Y may range from 500 to 1000 meters.

### *C. The proposed WSN-VANET data gathering and processing combined algorithm:*

Based on the above basic settings and assumptions, the proposed combined data gathering and processing algorithm (appropriate for large-scale environmental monitoring in rural areas) consists of the following steps (see also Fig. 1):

1. Each *SN* sends its measurements periodically to its *CH* through its parent node (intra-cluster communication).

2. Each *CH* gathers and sends the received measurements (together with its own) periodically to the corresponding *RN* (the *RN* of the relevant sub-area) through its parent *CH* node (inter-cluster communication).

3. Each *RN* gathers the received measurements and then it acts as follows:

    i. If it is close enough to an RSU (i.e. if an RSU resides in its range) it sends the received measurements directly to that RSU.

    ii. Else (if there isn't any RSU in its range) it broadcasts periodically an advertise message (*ADV_MSG*), which contains the next set of its gathered measurements.

4. Each vehicle *Vhi* (provided that is registered in the specific application framework) receiving the *ADV_MSG*, checks its ability flag (*fa*), which is set to '1' if the corresponding cluster (in which *Vhi* belongs) can still act as a virtual edge server/VEC provider (otherwise it's set to '0'). If *fa*=1 it sends an admittance message (*ADM_MSG*) to the advertising *RN*, setting itself as a candidate VEC-provider candidate for the next set of measurements.[1]

5. The advertising *RN* receives the *ADM_MSG* messages from the candidate VEC-providers, and sends them the next set of measurements in a first-come-first-serve (FCFS) basis.

6. Each vehicle *Vhj* that receives a set of measurements, then acts as follows (trying to schedule the preparation of as much intermediate results as possible, before forwarding data to the next RSU in the road/highway):

   i. It first defines the separate/independent tasks that can be executed using the newly received set of measurements, with respect to the specific application it's registered to – i.e. the air quality index – AQI – computation discussed in the next section as an experimental case).

   ii. It then looks for vehicles (cluster members – CMs) in its cluster, that may serve as probable VEC servers for the execution of one or more tasks, following the generalized algorithmic approach given in [30]. Moreover, it distinguishes between tasks with specific deadlines (tasks it will gather and aggregate the results itself before propagating them the next RSU), and tasks with no practical deadline, as indicated in more details in [30].[2]

   iii. Next, it sends the necessary data for the next task/s to each candidate *CM* that has responded positively and fulfills the requirements, in a first-come-first-serve (FCFS) basis.

   iv. If there are still remaining (not delivered) tasks, it first keeps one or more of them to execute locally (depending on its current load and its available resources), and it propagates the rest to the next RSU, when it falls in its range.

   v. Each vehicle *Vhj* that has received a task to execute, when it completes its execution, either (a) returns the results to the source vehicle *Vhi*, or (b) it keeps the results and sends them to the next RSU, when it falls in its range (the latter is necessary in case *Vhj*, *Vhi* have lost their connection to each other – i.e. one of them has lost its connection to the cluster).

   vi. Finally, the source vehicle *Vhi*, when it reaches the next RSU, except sending the probable remaining tasks, it also forwards to the corresponding RSU MEC-server all the intermediate results computed through the distributed edge processing approach described above.

---

[1] Note here, that the *CH* node of each cluster in the VANET is responsible to update (and broadcast to all the other vehicles / cluster members) the flag *fa* in tactical time intervals. The CH node is kept informed adequately each time a task is assigned to a *CM* through the corresponding finding process initiated by the source node. So, it can easily maintain a reliable estimation of the total task execution load of all the *CMs* and compute flag *fa* based also on the total available computing resources of the nodes.

[2] Note here, that in most cases (i.e. relevant applications – e.g. with regard to the AQI computation etc.), the corresponding tasks should not have specific/strict deadlines.

## 4. EXPERIMENTAL EVALUATION

In this section, our extended simulation experiments are presented, in which the proposed approach is shown to achieve high efficiency and scalability. Also, our approach is compared to the alternative approach of using only the WSN structure (along with the RSUs of the crossing road/highway), in order to show the worth of using the combined WSN-VANET scheme for large-scale environmental monitoring. Furthermore, we have selected as the basic environmental measurement case in our experiments, the computation of the well-known air quality index (AQI) value.

### 4.1 The Air Quality Index

The Air Quality Index (AQI) is an indicator introduced by government agencies to keep informed the public on how polluted the air currently is or how polluted it is forecasted to become [31]. Air quality is directly influenced by the increase in the concentration of pollutants. The Air Quality Index represents the severity of pollution for ordinary people. A lot of specific groups of individuals (children, the elderly, individuals with respiratory or cardiovascular problems etc.) are significantly affected by poor air quality. When the AQI is high, the central authorities generally take further actions to protect the people (i.e. encourage them to reduce outdoors physical activity, avoid going out for long intervals etc.).

Currently, there are two AQI indicators, one introduced by the US government (US-EPA AQI) [32], and one formed by the Indian government (CPCB AQI) [33]. Here we follow the first one, which has been extensively used as a general rule to measure contamination. Both the above AQI indicators adopt the same scale ranging from 0 (very good air quality) to 500 (very poor air quality), with the Indian index being more tolerant in cases of highly polluted air. A color code of AQI has also been developed for better representation of the results, aiming at giving a direct understanding of the air quality with a quick glance [34]. More specifically, for the computation of the AQI, six air pollutants have to be estimated: particulate matter (PM2.5 and PM10), carbon monoxide (CO), ozone (O3), nitrogen dioxide (NO2) and sulfur dioxide (SO2). The measured values are then converted to the AQI scale (0–500) as described in [34]. The AQI is finally determined by the highest of the above converted values. For the correct computation, measurements for at least three pollutants are required, in which one should be the number of particles of either PM2.5 or PM10 [35,36]. The formal expression used for the calculation of AQI for each element is the following [37]:

$$I_p = \frac{I_{Hi} - I_{Lo}}{BP_{Hi} - BP_{Lo}}(C_p - BP_{Lo}) + I_{Lo}$$

where $I_p$ is the value of pollutant p, $C_p$ is the concentration of pollutant p, $BP_{Lo}$ is the concentration breakpoint that is less than or equal to $C_p$, $BP_{Hi}$ is the concentration breakpoint that is greater than or equal to $C_p$, $I_{Lo}$ is the AQI value corresponding to $BP_{Lo}$ and $I_{Hi}$ is the AQI value corresponding to $BP_{Hi}$. For example, if we want to calculate the $I_p$ of PM10 particles with a concentration of $C_p$=162 μg/m3, then based on [34] we have $BP_{Lo}$=155, $BP_{Hi}$=254, $I_{Lo}$=101 and $I_{Hi}$=150, and using these values in the above formula, we get $I_p$=104.46. As also referred above, for the computation of AQI at least the measurement of three air pollutants is needed. So, if we follow this rule, and the measured values for each of the three pollutants are $I_{p1}$ = 104.46, $I_{p2}$ = 132.54 and $I_{p3}$ = 115.12, then $I_{p2}$ is the index with the highest value and the one that determines AQI.

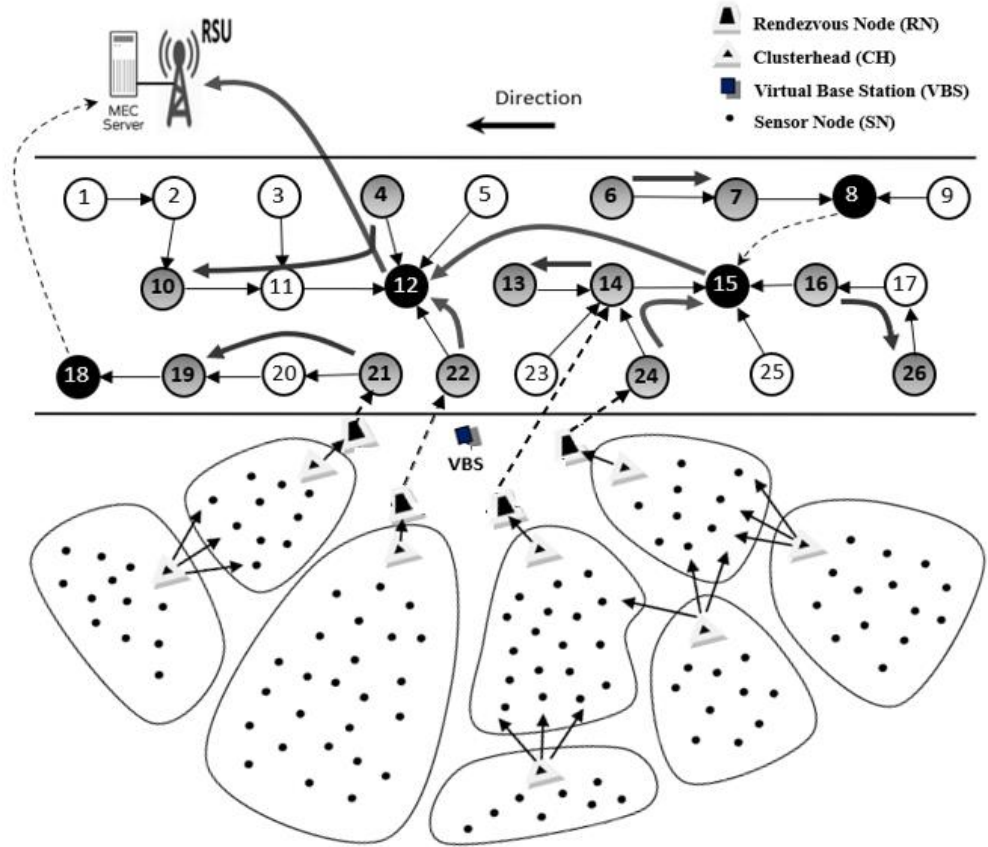

**Fig 1. A typical instance of the proposed scheme. Vehicles 21, 22, 14, 24 receive gathered sensed data from RNs, and either they offload tasks to other vehicles (21 to 19 and 14 to 13) or they forward the received data to the closest RSU.**

## 4.2 Experimental setup
Our experimental evaluation is based on extended simulations using Veins [38], SUMO [39] and Castalia [40] simulation tools as follows. With respect to the VANET structure we've used Veins which is based on OMNeT++, and includes a full stack of simulation models for investigating cars and infrastructure communicating via IEEE 802.11 based technologies. In Veins, the traffic simulation is performed by SUMO, a road traffic simulation package which can import city maps from a variety of file formats and support efficient simulations of very large networks with roads having multiple lanes, and also including simple moving (e.g. right-of-way) rules or traffic lights etc. More concretely, in our experiments, we assume that we have a three-lanes road/highway in both directions, of total length 100km, with eNodeB/RSU and MEC server support every X kms, where X ranges – in our experiments – from 1 km to 10 kms).

With respect to the WSN structure we've used the Castalia simulator, which is also based on OMNeT++. Specifically, we have run experiments for varying number of 'virtual' base-station points, each corresponding to a relevant SNs deployment sub-area). The extend of each such sub-area across the side of the road/highway is specified by the limits +Y/2 and -Y/2 meters (left and right respectively) from each virtual base-station point, where Y ranges – in our experiments – from 250 to 2500 meters. The depth of each sub-area is defined in 1km. In each such sub-area a number of approximately Z SNs are finally residing in average, where Z ranges – in our experiments – from 200 to 2000, depending on parameter Y, as well as on a specific density factor $f$. Each SN has a maximum transmission range R equal to 45m and its initial energy is equal to 500 Joules. The energy consumption for every transmission is based on the target distance and ranges from 29.04mW to 57.42mW (4.3m-45m). Finally, the energy consumption for reception is equal to 62mW, whereas for sleep mode it's equal to 0.016mW.

To achieve the collaborative use of both the above simulation environments (Castalia and Veins/SUMO), we've constructed a suitable script to feed the Veins components with the output values of the WSN simulation scenarios implemented in Castalia, in a periodic basis.

## 4.3 Experimental measurements and discussion
First, the high efficiency of the proposed WSN-VANET approach is measured, by comparing its performance to the alternative approach of using only the WSN structure. The alternative WSN-only approach implies that the RNs don't communicate with the vehicles/VANET at all; instead, they forward their gathered data to the neighboring RNs, and so on, till the closest RSUs of the crossing road/highway is met. More concretely, in Fig. 2, the completion time of the computation of the AQI (average value) is shown for both the above cases (WSN-VANET combine framework and WSN-only approach), based on one period sensing values through the entire WSN deployment area. The measurements used in Fig. 2 have been taken for a fixed value of Y=500m (i.e. a 500x1000m$^2$ sub-area), a fixed value of Z=400 SNs in each sub-area, and having the value of X ranging from 1 km to 10 kms (in the full-length road/highway of 100 Kms).

As it is shown in Fig. 2, a significant improvement (decrease in the total completion time) is achieved, when the combined WSN-VANET approach is used, which ranges from 5.8% (for X=1km) to 16.7% (for X=10km). Moreover, the reader may observe that the above improvement is getting larger as X increases (i.e. for smaller number of RSUs in the road/highway). The latter is naturally expected since in case of less RSUs (sparsely located in the road), the overhead of transferring the gathered data only through the WSN's CHs/RNs is getting larger, and the time savings offered by the use of the VANET structure (for intermediate processing and propagating of the gathered data) is more clear since it is used for greater portions of time.[3]

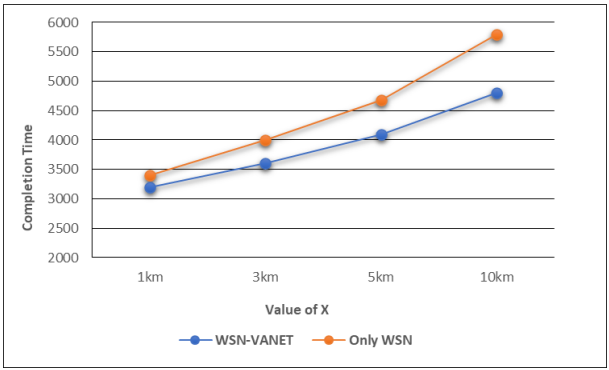

**Fig 2. Completion time with and without VANET**

Next (Fig. 3), the performance of the proposed scheme for different values of Y (i.e. for different sizes of the deployment sub-areas) is measured, having also the value of X ranging from 1km to 10kms (in the full-length road-highway of 100kms). More concretely, the corresponding measurements have been taken for Y ranging from 250m (i.e. a 250x1000m$^2$ sub-area with Z=200 SNs) to 2500m (i.e. a 2500x1000m$^2$ sub-area with Z=2000 SNs).

As it can be observed in Fig. 3, the total completion time increases as the value of Y increases, as it was expected since in that case the size of the deployment sub-areas is larger, the total number of RNs is less and their distribution becomes less canonical. Moreover, the corresponding increase is sharper for greater values of X, and the worst case is met for Y=2500m and X=10km, where the total completion time almost doubles over the most normal cases with smaller X and Y values. On the other hand, for Y=250-1000m and X=3-5km (values closer to the real word), the total completion time decreases a lot and it's actually very satisfactory.

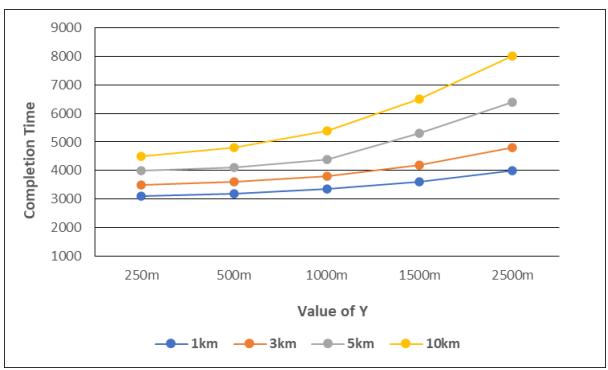

**Fig 3. Completion time for different values of Y**

---
[3] Note here that in real-world situations, the value of X is usually equal to 3-5kms; the achieved improvement in these cases is around 10%.

Finally (in Fig. 4), the scalability of our approach (as the length of the road/highway increases) is measured, for fixed value of Y=500m, and having the value of X ranging again from 1km to 10kms. More concretely, the corresponding measurements have been taken for the length of the road ranging from 25km to 100km.

As it can be observed in Fig. 4, the total completion time increases slowly with the increase of the road/highway length (especially in the most normal/real-world cases of X=1-5kms), which shows a very satisfactory scalability in large-scale deployment areas. The scalability factor gets worse only in the case of X=10km, which is however a quite weak case (not usually met in the real world) in terms of eNodeB/RSU coverage, especially in national roads/highways.

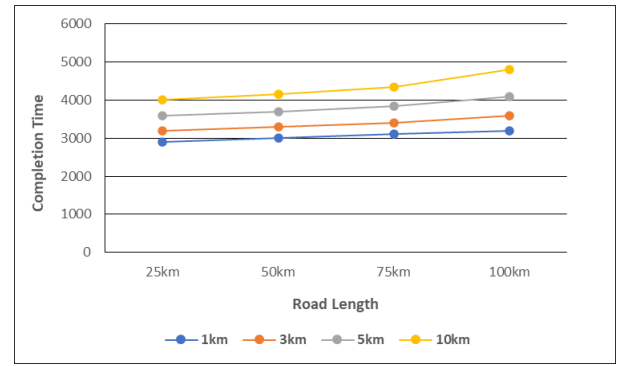

**Fig 4. Completion time for varying road length**

## 5. CONCLUSION
The worth of using WSNs and VANETs in collaboration, for the efficient gathering and processing of environmental parameters in long eNodeB/RSU-enabled roads/highways which cross large rural areas, is presented and discussed throughout this paper. The former (WSNs) is used for traditional sensing and data gathering, whereas the latter (VANET) is used for both performing intermediate processing and propagating the gathered data to the residing RSUs. Efficient node clustering and edge computation offloading techniques are elaborated to improve the behavior of the proposed scheme. Further, extended simulation experiments have been conducted (using SUMO, Veins and Castalia simulation platforms), which demonstrate the high efficiency of the proposed approach in the experimental field of air quality monitoring.